\documentclass[%
 reprint,
 amsmath,amssymb,
 aps,
prb,
noeprint,
]{revtex4-2}

\usepackage{graphicx}
\usepackage{bm}
\usepackage{hyperref}

\usepackage{xcolor}
\usepackage{physics}

\begin{document}


\title{Quantum-dot-based Kitaev chains: Majorana quality measures and scaling with increasing chain length}

\author{Viktor Svensson}
\affiliation{%
Division of Solid State
Physics and NanoLund, Lund University, S-221 00 Lund, Sweden
}
 \author{Martin Leijnse}
 \affiliation{%
Division of Solid State
Physics and NanoLund, Lund University, S-221 00 Lund, Sweden
}

\date{\today}

\begin{abstract}
Realizing Majorana bound states (MBSs) in short, well-controllable chains of coupled quantum dots sidesteps the problem of disorder, but requires fine-tuning and does not give the true topological protection inherent to long chains. Here, we introduce a new quality measure that is applicable also in the presence of strong electron-electron interactions and that quantifies the closeness to topological protection of finetuned MBSs in short quantum-dot chains. We call this measure local distinguishability because it puts a bound to the degree an arbitrary local measurement can distinguish between two  states. We study the local distinguishability for quantum-dot chains of different length. The three-dot chain is studied in detail, and we find that it may not always be an improvement over the two-dot case, a fact that can be understood within an effective model derived from perturbation theory. For longer chains, the local distinguishability vanishes exponentially, signalling a transition to a topological phase with two ground states that cannot be distinguished by any local measurement. 
\end{abstract}

\maketitle

\section{Introduction}\label{sec:introduction}
The past 15 years have seen a surge of efforts aiming to realize the physics of the so-called Kitaev chain \cite{kitaevUnpairedMajoranaFermions2001}, which features a topological superconducting phase. This phase is associated with Majorana bound states (MBSs), zero energy excitations with non-abelian statistics \cite{aliceaNonAbelianStatisticsTopological2011, beenakkerSearchNonAbelianMajorana2020, aliceaNewDirectionsPursuit2012, leijnseIntroductionTopologicalSuperconductivity2012, aguadoMajoranaQuasiparticlesCondensed2017} that could be used for topological quantum computation \cite{nayakNonAbelianAnyonsTopological2008, sarmaMajoranaZeroModes2015, marraMajoranaNanowiresTopological2022a}. 
Many different systems can be engineered to have a low-energy subspace that realizes an effective Kitaev chain, including, for example, proximitized nanowires \cite{oregHelicalLiquidsMajorana2010, lutchynMajoranaFermionsTopological2010, vaitiekenasFluxinducedTopologicalSuperconductivity2020, mourikSignaturesMajoranaFermions2012,dasZerobiasPeaksSplitting2012, dengMajoranaBoundState2016}, magnetic adatoms \cite{nadj-pergeProposalRealizingMajorana2013, nadj-pergeObservationMajoranaFermions2014}, and planar Josephson junctions \cite{pientkaTopologicalSuperconductivityPlanar2017, hellTwoDimensionalPlatformNetworks2017, fornieriEvidenceTopologicalSuperconductivity2019, renTopologicalSuperconductivityPhasecontrolled2019}. However, the disorder present in all these platforms complicates the realization and experimental detection of a topological superconducting phase and MBSs \cite{kellsZeroenergyEndStates2012, pradaTransportSpectroscopyNS2012,  liuZeroBiasPeaksTunneling2012, royTopologicallyTrivialZerobias2013, liuAndreevBoundStates2017, mooreQuantizedZerobiasConductance2018, reegZeroenergyAndreevBound2018, awogaSupercurrentDetectionTopologically2019, vuikReproducingTopologicalProperties2019, panPhysicalMechanismsZerobias2020, pradaAndreevMajoranaBound2020, hessLocalNonlocalQuantum2021}.

Because of this, the proposal to implement a Kitaev chain in well-controllable coupled quantum dots \cite{sauRealizingRobustPractical2012, leijnseParityQubitsPoor2012, fulgaAdaptiveTuningMajorana2013} has attracted increasing attention. Device complexity increases with more dots and short chains are preferable, but fortunately already the two-dot chain can host MBSs \cite{leijnseParityQubitsPoor2012}. The price to pay in short chains is that there is no true topological phase -- the system must be fine-tuned to a \textit{sweet spot} to show signatures of MBSs and some protection against perturbations. Because of the lack of formal topological protection, the associated states have been called poor man's MBSs. Various new ideas for how to design and tune such two-dot chains have been explored theoretically  \cite{liuTunableSuperconductingCoupling2022, tsintzisCreatingDetectingPoor2022, liuEnhancingExcitationGap2024, samuelsonMinimalQuantumDot2024, lunaFluxtunableKitaevChain2024}, and experimentally realized \cite{wangSingletTripletCooper2022, dvirRealizationMinimalKitaev2023, zatelliRobustPoorMans2024, tenhaafTwositeKitaevChain2024}. 

Experimentally distinguishing MBSs from topologically trivial Andreev bound states has proven to be a difficult problem. Experiments on the quantum-dot platform \cite{dvirRealizationMinimalKitaev2023, zatelliRobustPoorMans2024, tenhaafTwositeKitaevChain2024} have already demonstrated correlated appearance and splittings of zero-bias peaks at both ends of the chain, as well as signatures in the nonlocal conductance that are consistent with (poor man's) MBSs. Conductance measurements in a setup including an additional probe quantum dot could provide additional estimates of the nonlocal properties of the wavefunction \cite{soutoProbingMajoranaLocalization2023}. However, making conclusive statement about the closeness of the observed states to true topological MBSs would require protocols aimed at demonstrating non-abelian physics \cite{liuFusionProtocolMajorana2023, borossBraidingbasedQuantumControl2024, tsintzisMajoranaQubitsNonAbelian2024}. It is also an interesting theoretical problem to define a MBS quality measure, that quantifies the similarity of a given state to a true topological MBS. One such measure that is based on the MBS wavefunction is the Majorana polarization (MP) \cite{sedlmayrVisualisingMajoranaBound2015, sedlmayrMajoranaBoundStates2016, aksenovStrongCoulombInteractions2020, tsintzisCreatingDetectingPoor2022}.  
However, in the presence of electron-electron interactions, which are typically very strong in the quantum-dot platform, the many-body nature of the MBSs \cite{stoudenmireInteractionEffectsTopological2011, wrightLocalizedManyParticleMajorana2013, obrienManyparticleMajoranaBound2015, obrienManybodyInterpretationMajorana2015, kellsManybodyMajoranaOperators2015,  kellsMultiparticleContentMajorana2015a, bozkurtInteractioninducedStrongZero2024} complicates the interpretation of the MP. 

In this article, we introduce a MBS quality measure based on the local distinguishability of ground states in topological phases, which has no extra difficulty in interacting systems and gives a rigorous bound on the protection against perturbations. Increasing the length of the system should produce better quality MBSs, and recent experimental results on a three-dot Kitaev chain \cite{bordinCrossedAndreevReflection2023, bordinSignaturesMajoranaProtection2024} shows an increased stability of the energy spectrum under certain perturbations. There is also recent theoretical work on scaling up to longer length chains \cite{liuProtocolScalingSignordered2024, ezawaEvenoddEffectRobustness2024}. Here, we analyze the protection as a function of the chain length, paying particular attention to the three-dot case. We find that the protection in this case may be \textit{worse} in some respects than the two-dot case, and we explain these results with an effective Hamiltonian derived by perturbation theory. 

The paper is organized as follows: Section \ref{sec:system} describes the system and Section \ref{sec:protection_types} discusses how to characterize and quantify the quality of MBSs and how to tune a system to a sweet spot in parameter space where such quality measures are maximized. Section \ref{sec:three-dots} then focuses on how to tune the three-dot chain and shows that it can be worse than the two-dot chain in some respects, while Section \ref{sec:effective_model} contains a derivation of an effective model to explain these results. Finally, Section \ref{sec:scaling} shows what happens when scaling to longer chains. 

\section{System}\label{sec:system}
\begin{figure}[th!]
    \centering
    \includegraphics[width=\linewidth]{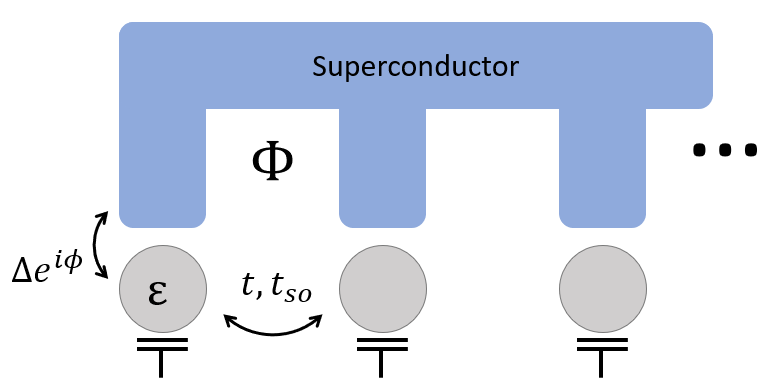}
    \caption{Sketch of setup where each quantum dot is coupled to the bulk superconductor via a local proximity effect (with amplitude $\Delta$ and phase $\phi$), and to its neighbours via both spin-preserving and -flipping tunneling (with amplitudes $t$ and $t_\text{so}$). The dot levels $\varepsilon$ and magnetic flux $\Phi$ (controlling the phase $\phi$) are used to tune to a sweet spot.}
    \label{fig:system}
\end{figure}

The system we consider consists of a chain of quantum dots, each locally coupled to a superconductor and coupled to each other via tunneling, see Fig.\,\ref{fig:system}. This setup was proposed in Ref.\,\cite{fulgaAdaptiveTuningMajorana2013} and the two-dot interacting version was studied in detail in Ref.\,\cite{samuelsonMinimalQuantumDot2024}, where it was shown that for a wide range of parameters it exhibits a sweet spot with localized MBSs, which can be reached by tuning the dot levels $\varepsilon_j$ and the superconducting phases $\phi_j$ (controlled by the magnetic fluxes $\Phi_j$ through the loops). As $V_Z, \Delta \rightarrow \infty$, the system reduces to a Kitaev chain.

The system Hamiltonian consists of two parts,
\begin{equation}
    H = H^\text{QD} + H^C,
\end{equation}
where $H^\text{QD}$ contains all the terms local to each of the $N$ dots and $H^C$ contains the coupling between them. These parts take the form
\begin{subequations}\label{eq:hamiltonians}
\begin{align}
    H^\text{QD} &= \sum_{j=1}^N \sum_{\sigma} (\varepsilon_j + \eta_\sigma V_{Z,j})n_{j\sigma} \nonumber\\
    &+ (\Delta_je^{i\phi_j}  d_{j\uparrow}^\dagger d_{j\downarrow}^\dagger + \text{h.c.}) + U_{l,j} n_{j\uparrow}n_{j\downarrow} \\
    H^C &= \sum_{j=1}^{N-1} \sum_{\sigma} (t_j d_{j\sigma}^\dagger  d_{j+1\sigma} + \text{h.c.}) \nonumber\\ 
    &+ ( \eta_\sigma t_{\mathrm{so},j} d_{j\sigma}^\dagger d_{j+1\bar{\sigma}} + \text{h.c.}) + U_{\mathrm{nl},j}N_j N_{j+1}, 
\end{align}
\end{subequations}
where $d^\dagger_{j\sigma}$ creates an electron with spin $\sigma$ on dot $j$, $n_{j\sigma} = d^\dagger_{j\sigma}d_{j\sigma}$, $N_j = n_{j\uparrow} + n_{j\downarrow}$ and $\eta_\sigma = \pm$. The local term includes the dot levels $\varepsilon_j$, Zeeman splitting $V_{Z}$, induced local pairing $\Delta_j$ and its phase $\phi_j$, and the local Coulomb interaction $U_{l,j}$. The term coupling dots together includes spin-conserving and spin-flip tunnelling $t_j$ and $t_{\mathrm{so},j}$, and the non-local Coulomb interaction $U_{\mathrm{nl},j}$.

\section{Protection and sweet spot quality}\label{sec:protection_types}
In this section, we discuss how to characterize the kind of protection implied by the existence of MBSs, the essential differences between short and long chains, and define how we search for a sweet spot. We denote by $\ket{o}$ and $\ket{e}$ the ground states in the odd and even fermionic parity sectors respectively and define $\delta\rho = \dyad{o}-\dyad{e}$,  Their energy difference is $\delta E$ and $E_\text{ex}$ is the gap to excited states. 

The existence of non-overlapping Majorana operators implies that the two ground states can't be distinguished by \textit{any} local measurement. Local means here that the spatial size of the terms in the operator is smaller than the separation between the MBSs. In particular, if the Hamiltonian is local, then $\delta E = 0$. This is the case for a long Kitaev chain in the topological phase. This degeneracy is stable under any local perturbation $\mathcal{O}$ as  
\begin{align}\label{eq:deltaE_tr}
    \delta E &= \expval{H+\mathcal{O}}_o - \expval{H+\mathcal{O}}_e \nonumber \\
    & = \Tr[\mathcal{O}\delta\rho]
\end{align}
corresponds to the difference of a local measurement of the states. Characterizing the quantity $\Tr[\mathcal{O}\delta\rho]$ will be one of the main points of this article.

In short chains, even non-overlapping MBSs may be close enough to be coupled together by the Hamiltonian. The Hamiltonian is in this sense \textit{non-local} and $\delta E$ may be non-zero even if the MBSs are perfectly localized to the edges. This splitting introduces an undesirable dynamical phase in fusion and braiding protocols. For a non-local Hamiltonian, $\delta E=0$ is an independent condition from the separation of MBSs. Therefore, as we search for the sweet spot, we will consider these independently.

\subsection{Protection from fluctuating parameters}
The stability of $\delta E$ is not necessarily related to the presence of separated MBSs. Consider a Hamiltonian $H(\lambda)$ that is a function of some parameter $\lambda$. Protection against fluctuations in $\lambda$ may arise in several ways, some of which are
\begin{enumerate}
    \item $\norm{\frac{\partial H}{\partial\lambda}\delta\lambda}$ is small because the Hamiltonian is naturally insensitive to variations in $\lambda$.
    \item $\Tr[\delta\rho \frac{\partial H}{\partial\lambda}]$ is small, because $\delta\rho$ is orthogonal to this particular perturbation.
    \item There are non-overlapping MBSs that map between the ground states and the Hamiltonian is local. 
\end{enumerate}
The first type has nothing to do with MBSs. The second type is implied by the third, but can also occur when the MBSs overlap, see Sec.\,\ref{sec:three-dot-homogeneous}. The third type is the strongest and implies protection based only on the \textit{locality} of the perturbations and the MBS wavefunction. 

\subsection{Majorana Polarization}\label{sec:MPP}
In non-interacting systems the MBS wavefunction is simple to define and the Majorana polarization (MP) can be used to characterize how separated the two MBSs are \cite{sedlmayrVisualisingMajoranaBound2015, sedlmayrMajoranaBoundStates2016}. A generalization has been used in interacting systems\cite{aksenovStrongCoulombInteractions2020, tsintzisCreatingDetectingPoor2022, samuelsonMinimalQuantumDot2024} but in this case it cannot fully characterize MBSs with non-trivial many-body content \cite{wrightLocalizedManyParticleMajorana2013, obrienManyparticleMajoranaBound2015, obrienManybodyInterpretationMajorana2015, kellsManybodyMajoranaOperators2015,  kellsMultiparticleContentMajorana2015a}. We use the definition from Ref.\,\cite{samuelsonMinimalQuantumDot2024},
\begin{equation}\label{eq:majoranapolarization}
    \mathrm{MP} = \frac{\abs{\sum_{l,s} \matrixelement{e}{\gamma_{l s}}{o}^2}}{\sum_{l,s} \abs{\matrixelement{e}{\gamma_{l s}}{o}^2}},
\end{equation}
where $\ket{o}$ and $\ket{e}$ are the odd and even ground states, $l$ indexes the fermions on the leftmost site \footnote{In the full model, there are two fermions on each site because there are two spins. In the effective model, see Sec.\,\ref{sec:effective_model}, there is only one.} and
\begin{subequations}
\begin{align}
    \gamma_{l +} &= d_{l} + d_{l}^\dagger, \\
    \gamma_{l -} &= i(d_{l} - d_{l}^\dagger).
\end{align}
\end{subequations}
We have chosen to focus on the outer dots, and by reflection symmetry, we don't need to consider the rightmost dot.
For a non-interacting system, $\mathrm{MP} = 1$ implies that there is no overlap of the MBSs on the outer dots and the system is protected from perturbations at those dots. However, with interactions, or imperfect $\mathrm{MP}$, there is no clear quantitative relation between the $\mathrm{MP}$ and protection from perturbations. 

\subsection{Local distinguishability}\label{sec:LD}
Due to the difficulties with the MBS wavefunction, we introduce a measure based on the local distinguishability of the ground states. This measure provides a rigorous bound on the stability of $\delta E$ and is well-defined even with interactions. Our construction is inspired by Ref.\,\cite{obrienManybodyInterpretationMajorana2015} where the presence of MBSs is diagnosed by probing the ground states with local measurements such as the charge on each dot. We extend this to cover $\textit{all}$ local measurements.

Consider a perturbation $\mathcal{O}_R$ acting only in some subsystem $R$. The rest of the system is denoted by the complement $R^\complement$. To first order in this perturbation, the energy difference between the ground states changes if the perturbation can distinguish the states, see Eq.\,\ref{eq:deltaE_tr}. This quantity obeys
\begin{equation}
    |\Tr[\mathcal{O}_R\delta\rho]| = |\Tr[\mathcal{O}_R\delta\rho_R]| \leq \norm{\mathcal{O}_R}\norm{\delta\rho_R}
\end{equation}
where $\delta \rho_R = \Tr_{R^\complement}\delta \rho$ is the difference in reduced density matrices and $\norm{\cdot}$ is the Frobenius norm \footnote{Different bounds result from using Hölder's inequality $|\Tr[\mathcal{O}_R\delta\rho]| \leq \norm{O_R}_q \norm{\delta\rho_R}_p$, where $1/q +1/p = 1$ and $\norm{\cdot}_q$ is the Schatten norm. Without additional assumptions on the perturbation, we have no reason to prefer one over another.}. We see that $\norm{\delta\rho_R}$ bounds the effect of any perturbation in the region $R$. In App.\,\ref{app:LD} we show that it is zero if there exists a MBS outside of the region $R$. 

In a topological phase, $\norm{\delta\rho_R} = 0$ in all sufficiently local regions. For our system, we take the subsystems to be each individual dot, and define the local distinguishability ($\mathrm{LD}$) as 
\begin{equation}
    \mathrm{LD} = \sqrt{\sum_i \norm{\delta\rho_{QD_i}}^2}
\end{equation}
which we minimize when optimizing for the sweet spot.

$\mathrm{LD}$ is simple to compute as it only relies on finding ground states. It can be calculated in large interacting systems by using e.g. DMRG. In non-interacting systems, the reduced density matrix can be reconstructed from the single particle density matrix. 

\subsection{Optimizing for a sweet spot} \label{sec:sweet_spot_definition}
In this section, we define how we optimize for sweet spots, both with and without energy degeneracy and introduce notation to keep the different cases apart. The term sweet spot is used interchangebly for all of them.

We denote by $\mathrm{mLD}_0$ the minimal LD point with degenerate ground states, i.e. the solution to the minimization problem
\begin{align}
\begin{split}    
\min \quad & \mathrm{LD}  \\
\text{s.t.} \quad & \delta E = 0.
\end{split}
\end{align}
In practice, we impose the latter condition by solving the problem
\begin{align}
\min \quad & \mathrm{LD} + \lambda \abs{\delta E}
\end{align}
which includes a penalty factor $\lambda$ that enforces the constraint. We will compare this point to the point denoted by $\mathrm{mLD}$, defined by the solution to
\begin{align*}
\min \quad & \mathrm{LD},
\end{align*}
which does not require the states to be degenerate.

Another approach to defining a sweet spot, used in e.g. Refs.\,\cite{lunaFluxtunableKitaevChain2024, bozkurtInteractioninducedStrongZero2024}, is to optimize for a spot with degenerate ground states and where that degeneracy is protected against fluctuations in the dot levels $\varepsilon$. In other words, where $\frac{\partial \delta E}{\partial \varepsilon} = 0$. We call this the level-protected point. Since we also tune the phase difference $\delta\phi$, we also consider a phase-protected point defined analogously. These points are in general different from the $\mathrm{mLD}$ points.

\section{Results}
We start by analyzing the three-dot case in Sec.\,\ref{sec:three-dots}, comparing the different sweet spots in detail. The observations will be explained with perturbation theory in Sec.\,\ref{sec:effective_model}, and in Sec\,\ref{sec:scaling} we consider longer chain lengths. Our results are produced with the code available at \cite{svenssonCvsvenssonLongerPoorMansMajoranas2024}.

The constant parameters are set to
\begin{subequations}
\begin{align}
t_\text{so}/t &= 5 \\
\sqrt{t^2+t_\text{so}^2} &= \Delta/2 \\
V_Z &= 3\Delta.
\end{align}
\end{subequations}
When $t>t_\text{so}$, it is advantageous to tune $\varepsilon$ in an alternating manner to effectively enhance the spin-flipping process \cite{samuelsonMinimalQuantumDot2024}. Picking $t<t_\text{so}$ is simpler, because the sweet spot is closer to homogeneous. Similar results are expected in both cases as the effective Hamiltonians (see Sec.\ref{sec:effective_model} and App.\,\ref{app:perturbations}) are the same at first order in perturbation theory.

For simplicity we also set
\begin{equation}
U_l = U_{\mathrm{nl}} = 0
\end{equation}
for the results in the main paper as we found that the essential features we focus on in this work are qualitatively the same with and without Coulomb interactions. The main thing to keep in mind is that there is a limit to how large interaction strengths can be tolerated, see App.\,\ref{app:interactions} for details.

\subsection{Three-dot system}\label{sec:three-dots}
In this section we consider a three-dot system and how to tune it to a sweet spot. Assuming reflection symmetry, we have three independent parameters to tune, $\varepsilon_1 =\varepsilon_3 = \varepsilon, \, \varepsilon_2, \, \delta\phi =\phi_{2}-\phi_{1}= \phi_{3}-\phi_{2}$. In Sec.\,\ref{sec:three-dot-homogeneous}, we reduce it to two parameters by setting $\varepsilon_1=\varepsilon_2$, making it easy to visualize. In Sec.\,\ref{sec:three-dot-detuning} we study how the sweet spot can be improved by tuning $\varepsilon_2$ independently.

\subsubsection{Homogeneous tuning}\label{sec:three-dot-homogeneous}
\begin{figure}[t!]
    \centering
    \includegraphics[width=1\linewidth]{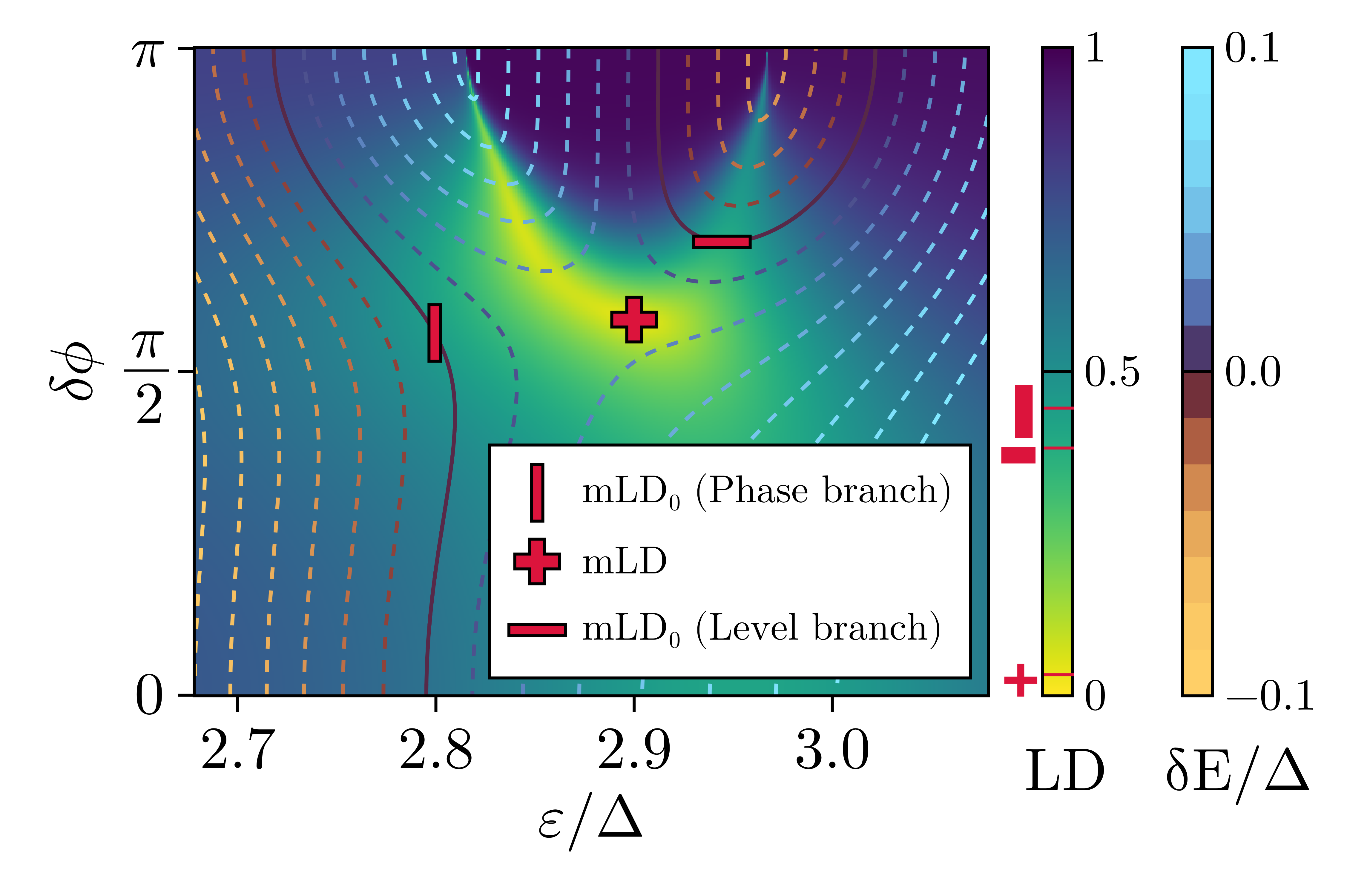}
    \caption{Tuning $\varepsilon$ and $\delta\phi$ homogeneously in the three-dot chain. Brighter color means a smaller $\mathrm{LD}$. The contours show the energy splitting between ground states $\delta E$, with solid line signifying degeneracy. The three points marked are defined in Sec\,\ref{sec:sweet_spot_definition}, and correspond to points with minimal $\mathrm{LD}$, with and without energy degeneracy.}
    \label{fig:tuning_sweet_spot}
\end{figure}

We start with a three-dot system where we tune the dot levels $\varepsilon$ and phase gradient $\delta\phi$ homogeneously. The heatmap in Fig.\,\ref{fig:tuning_sweet_spot} shows $\mathrm{LD}$ as these parameters are tuned, and the contour lines show the energy splitting between the two ground states. Unfortunately, and differently from the two-dot case, the solid lines where the ground states are degenerate do not lie in the region with well separated MBSs and good protection. 

In Fig.\,\ref{fig:tuning_sweet_spot}, the condition $\delta E = 0$ is satisfied on the solid contour lines, which has two disconnected branches. One branch features a point where it the contour is tangent to the horizontal axis. At this point, the degeneracy is protected from variations in $\varepsilon$. We call this the level branch. The other branch is called the phase branch as it features a point tangent to the vertical axis with protection against phase fluctuations. These tangential points feature protection against a specific perturbation, an example of type 2 protection as defined in Sec.\,\ref{sec:protection_types}.

On each branch, we can also define a point with minimal $\mathrm{LD}$ and $\delta E=0$, denoted by $\mathrm{mLD}_0$ and marked in the figure. These differ from the tangential points, but on the level branch they are very close. The constraint of $\delta E=0$ severely limits the quality of the sweet spot. The point marked $+$ and denoted by $\mathrm{mLD}$ has a non-zero $\delta E$ but $\mathrm{LD}$ is much smaller. 
Here, there are well-separated MBSs and the energy splitting is to some degree protected against \textit{any} local perturbation, but the ground states are not degenerate. We will see why in Sec.\,\ref{sec:effective_model}.

\subsubsection{Inhomogeneous tuning} \label{sec:three-dot-detuning}
\begin{figure}[t!]
    \centering
    \includegraphics[width=1\linewidth]{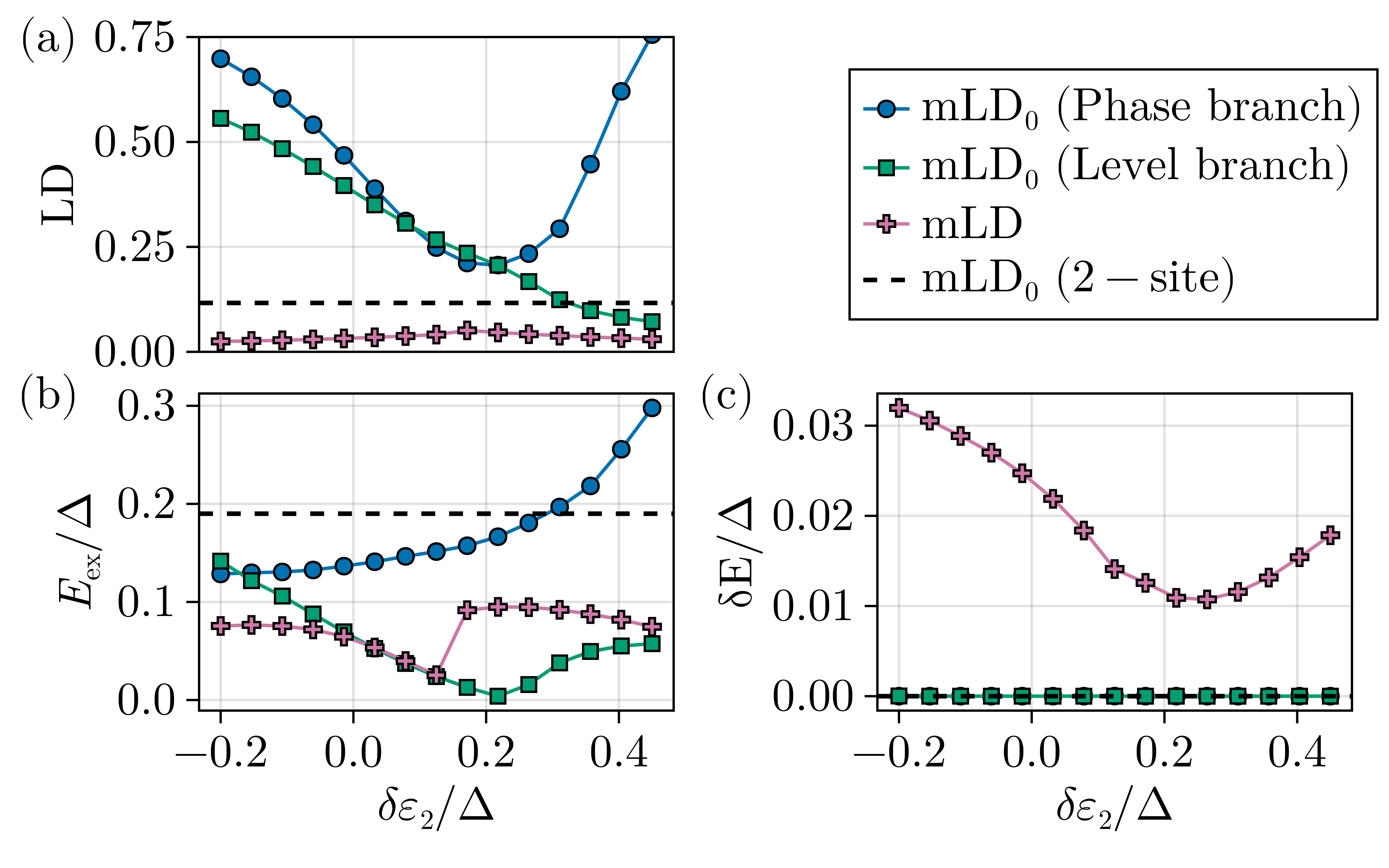}
    \caption{Sweet spot properties as a function of the middle dot detuning $\delta\varepsilon_2$. The $\mathrm{mLD}_0$ points on the two branches improve, but can't clearly beat the two-dot sweet spot without reducing the excitation gap. As the detuning increases, the phase branch becomes irrelevant while the level branch should be thought of as an effective two-dot chain were the middle dot acts as a tunnel barrier. At $\delta\varepsilon_2 \approx \Delta/5$, the location of the $\mathrm{mLD}$ point changes location discontinuously, see App.\,\ref{app:branches}.}
    \label{fig:3site_inhomogeneous}
\end{figure}

Fig.\,\ref{fig:tuning_sweet_spot} shows that there is a tension between energy degeneracy and good MBSs at the sweet spot. Here, we check that this is still the case even if the system is tuned inhomogeneously. We detune the middle dot so that $\varepsilon_2 = \varepsilon + \delta\varepsilon_2$.

In Fig.\,\ref{fig:3site_inhomogeneous}, we follow the phase branch and level branch of the sweet spot as a function of the detuning $\delta\varepsilon_2$. The sweet spots improve, especially the phase branch which reaches an $\mathrm{LD}$ and excitation gap comparable to the two-dot case. When the detuning gets large enough, the phase branch quickly gets worse, while the level branch features a small $\mathrm{LD}$ but with a bad excitation gap. Effectively, the middle dot has been detuned to act as a tunnel barrier and the three-dot chain can be effectively reduced to a two-dot chain. We conclude that the three-dot chain does not improve on the two-dot chain in this way of measuring the protection. Note that our definition of $\mathrm{LD}$ does not cover all possible perturbations, only those local to each dot. It also only quantifies first order protection, and says nothing about higher order terms.

\subsection{Effective model}\label{sec:effective_model}
The three-dot chain seems to offer a \textit{worse} protection against local perturbations than the two-dot chain when we impose energy degeneracy. In this section we show why, by deriving an effective model to second order in $t/V_Z$. The second order terms give a Kitaev chain that includes next-nearest neighbours hoppings and pairings. This makes the effective Hamiltonian less local, and it can then more easily distinguish the ground states.

\subsubsection{Deriving the effective model}\label{sec:effective_model_derivation}
We can derive an effective Kitaev model using perturbation theory. We follow \cite{fulgaAdaptiveTuningMajorana2013, samuelsonMinimalQuantumDot2024} and treat the hopping as a perturbation of $H^\text{QD}$. We first transform $d_n \rightarrow d_n e^{i\phi_n/2}$ to put the phase winding on the hopping. $H^\text{QD}$ is diagonalized by a Bogoliubov transformation
\begin{subequations}
\begin{align}
d_{n \downarrow }^{\dagger }&= \frac{
\sqrt{\beta_n-\epsilon_n} a_n^{\dagger }-\sqrt{\beta_n+\epsilon_n} b_n^{\dagger }
}{\sqrt{2\beta_n}} \\
d_{n \uparrow } &= \frac{\sqrt{\beta_n+\epsilon_n} a_n^{\dagger }+\sqrt{\beta_n-\epsilon_n} b_n^{\dagger }}{\sqrt{2\beta_n}}
\end{align}
\end{subequations}
where $\beta_j = \sqrt{\varepsilon_j^2 + \Delta_j^2}$. The $a$ and $b$ fermions have energies
\begin{align*}\label{eq:aenergy}
    E_{a_j} &= \beta_j - V_Z,\\
    E_{b_j} &= \beta_j + V_Z,
\end{align*}
and we consider the regime where $E_a \ll E_b$ and integrate out the $b$-fermions. Let $P$ be the projector on the subspace with no occupied $b$-fermions, and $Q = I - P$. To second order in the hoppings, the effective Hamiltonian in the subspace $P$ is
\begin{align}
    H^\text{eff} &= PHP - \frac{1}{2E_b}PH^\text{C} Q H^\text{C}P  \nonumber \\
     &= H^\text{eff}_1 + H^\text{eff}_2.
\end{align}
The $b$-fermions have been integrated out, but leave a trace in the form of the second order term. 

The first term is a standard Kitaev chain, 
\begin{equation}
    H_1^\text{eff} = \sum_n \varepsilon_{a} a_n^\dagger a_n + (t_{aa} a_n^{\dagger } a_{n+1} +\Delta_{aa} a_n a_{n+1} + \mathrm{h.c})
\end{equation}
as derived previously for this model in \cite{fulgaAdaptiveTuningMajorana2013, samuelsonMinimalQuantumDot2024}. The explicit expressions for the parameters for a homogeneous system are shown in App.\,\ref{app:perturbations}.
The second term,  
\begin{multline}
    H_2^\text{eff} = \sum_n \varepsilon_{2} a_n^\dagger a_n + t_{2} a_n^\dagger a_{n+2} + \Delta_{2} a^\dagger_{n}a^\dagger_{n+2} + \mathrm{h.c},
\end{multline}
includes longer range terms that couple next nearest neighbours. This is not unique to this system, the same thing happens in other dot-based Kiteav chains \cite{milesKitaevChainAlternating2024}. Kitaev chains with long range couplings have been studied in Refs.\,\cite{degottardiMajoranaFermionsSuperconducting2013, alecceExtendedKitaevChain2017, mahyaehZeroModesKitaev2018}. These terms are to blame for the complications of the three-dot sweet spot, as they make the Hamiltonian more non-local.

\begin{figure}[t!]
    \centering
    \includegraphics[width=1\linewidth]{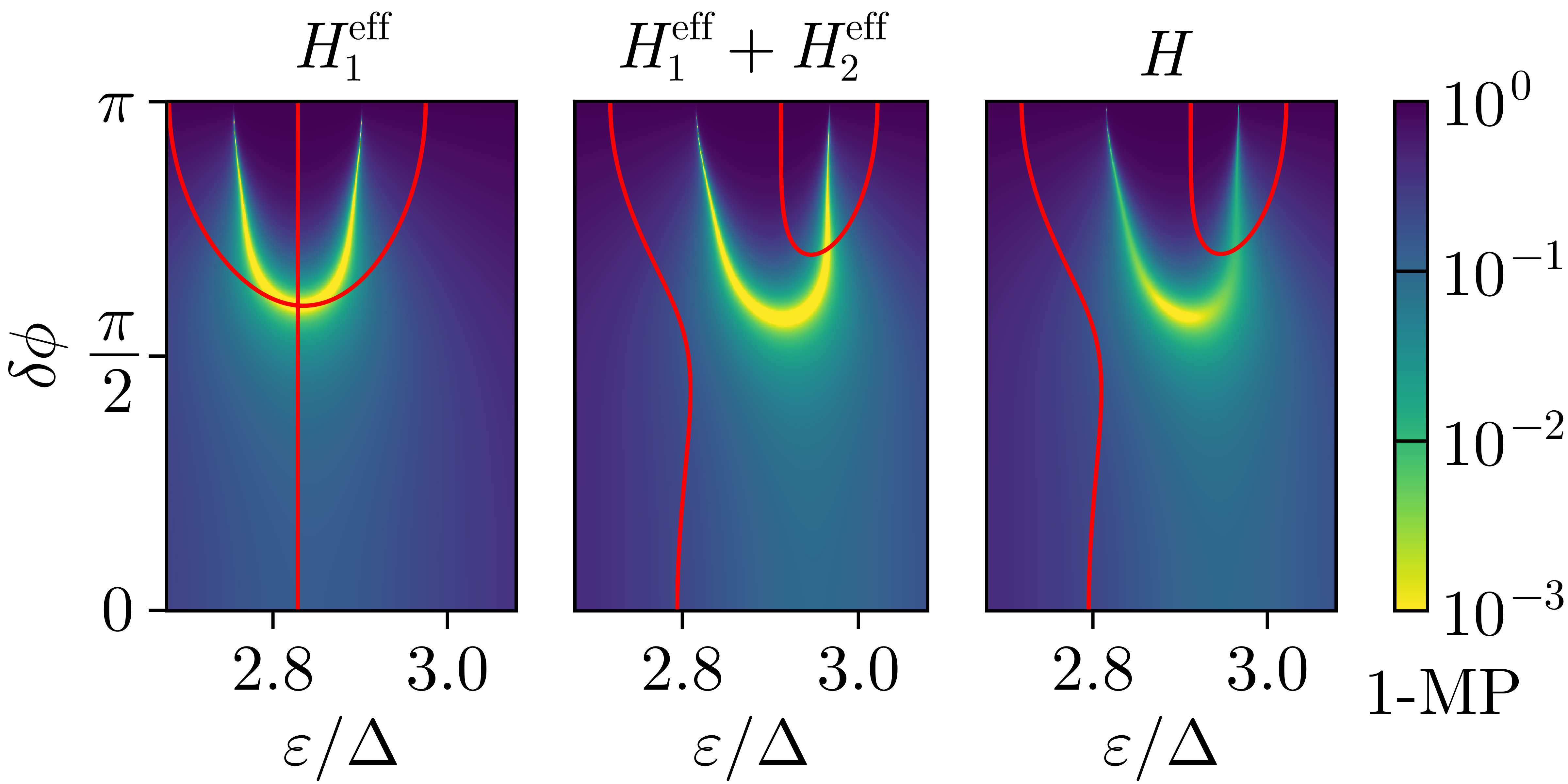}
    \caption{The heatmap shows the $\mathrm{MP}$ (see Eq.\,\ref{eq:majoranapolarization}) and the red contour shows where the ground states are degenerate. The first order effective Hamiltonian has a sweet spot where $\mathrm{MP}$ is maximized and the two branches of $\delta E=0$ coincide. In the second order model (which closely resembes the full model), the $\delta E = 0$ lines does not coincide with the maximum in $\mathrm{MP}$. This is due to the additional long range couplings which can be significant if the Zeeman splitting is finite.}
    \label{fig:perturbative_tuning}
\end{figure}

We confirm this with Fig.\,\ref{fig:perturbative_tuning}. It shows the tuning plot, now for the $\mathrm{MP}$ for the first and second order effective Hamiltonian, and for the full model. The first order effective model, which corresponds to a standard Kitaev chain, has a definitive sweet spot where the ground states are degenerate, $\mathrm{MP} = 1$ and $\mathrm{LD} = 0$. The two branches where $\delta E = 0$ cross at the sweet spot and the contours are tangential to both axes. When the longer range terms of $H^\text{eff}_{2}$ are included, the plot closely resembles the result of the full model, where the branches avoid each other as well as the point with maximally separated MBSs. Comparing Figs.\,\ref{fig:tuning_sweet_spot} and \ref{fig:perturbative_tuning} we also see that the two measures of sweet spot quality, $\mathrm{MP}$ and $\mathrm{LD}$, are correlated.

\subsection{Scaling to longer chains}\label{sec:scaling}
In this section, we optimize homogeneously over dot levels and phase differences to find the sweet spot as a function of the length of the chain. At large system sizes, exact diagonalization of the many body Hamiltonian is prohibitively expensive, but since we consider a non-interacting theory at this point, the reduced density matrices $\rho_{\text{QD}_i}$ on each dot can be reconstructed from the correlators $\expval*{c_i^\dagger c_j}$ and $\expval*{c_i^\dagger c_j^\dagger}$\cite{chungDensitymatrixSpectraSolvable2001, cheongManybodyDensityMatrices2004} which are obtainable via standard Bogoliubov-de-Gennes techniques. 

\begin{figure}[t!]
    \centering
    \includegraphics[width=1\linewidth]{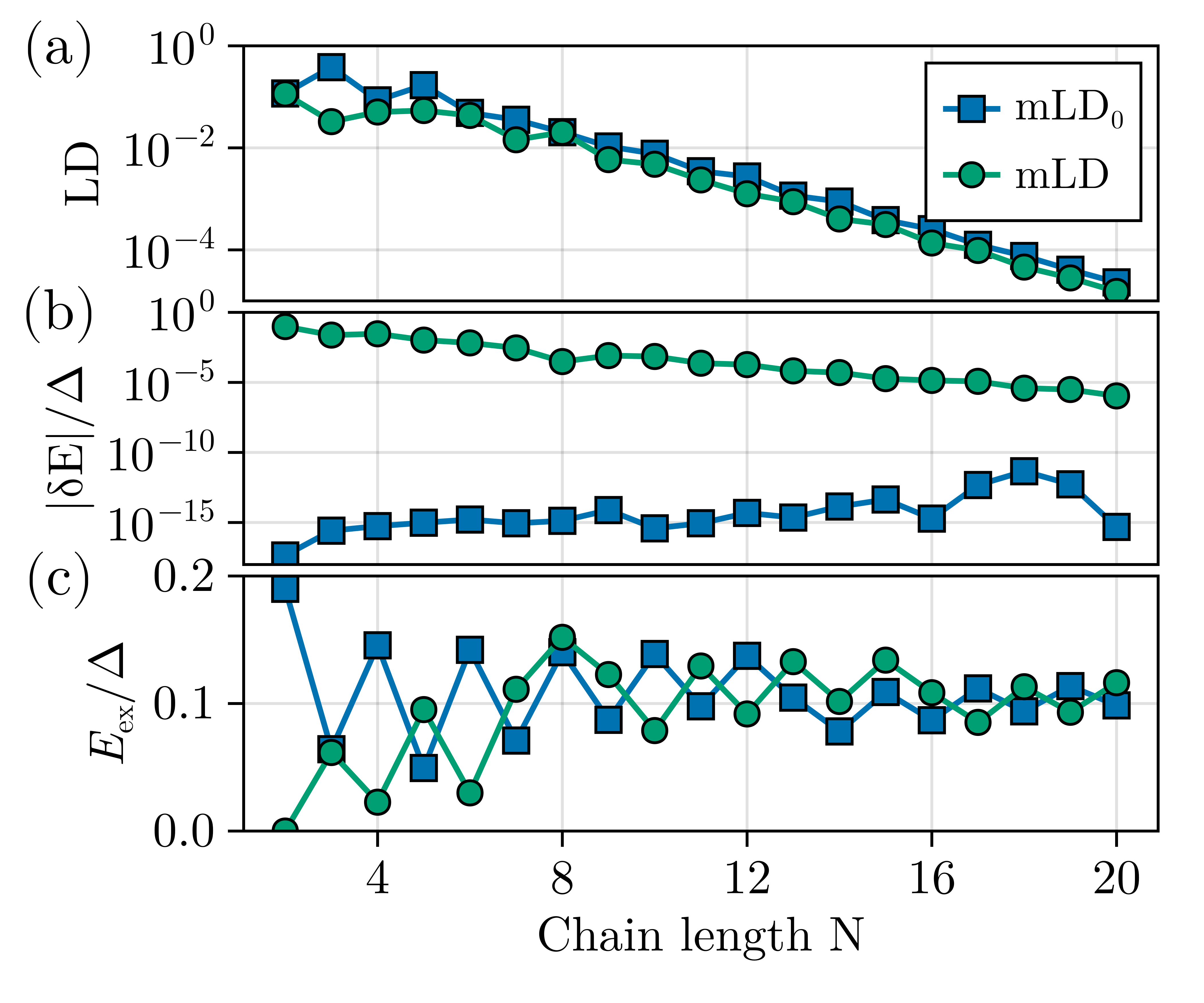}
    \caption{Sweet spot quality as a function of chain length with homogeneous tuning. (a) $\mathrm{LD}$ has an exponential fall-off at large chain lengths, implying exponentially better protection. At small lengths, the pattern is less clear. (b) The energy splitting $\delta E$. (c) The excitation gap is somewhat stable around $E_\text{ex} \sim \Delta/10$ for long chains. For $N=2$ and $4$, the excitation gap at $\mathrm{mLD}$ is very small, with only a very small improvement in $\mathrm{LD}$ over $\mathrm{mLD}_0$.}
    \label{fig:LD_scaling}
\end{figure}

As seen in Fig.\,\ref{fig:LD_scaling}(a), $\mathrm{LD}$ at the sweet spot falls off exponentially as the chain length increases. This is expected in a topological phase. Fig.\,\ref{fig:LD_scaling}(b) shows the splitting of the ground state degeneracy $\delta E$, which is constrained to be very small for $\mathrm{mLD}_0$. At the point $\mathrm{mLD}$, there is no constraint on the energy splitting, but since $\mathrm{LD}$ falls off exponentially, one would expect $\delta E$ to follow the same pattern. Fig.\,\ref{fig:LD_scaling}(c) shows the excitation gap at the sweet spot, which has an odd-even pattern but is otherwise quite stable at $E_\text{ex} \sim \Delta/10$. 

\begin{figure}[t!]
    \centering
    \includegraphics[width=1\linewidth]{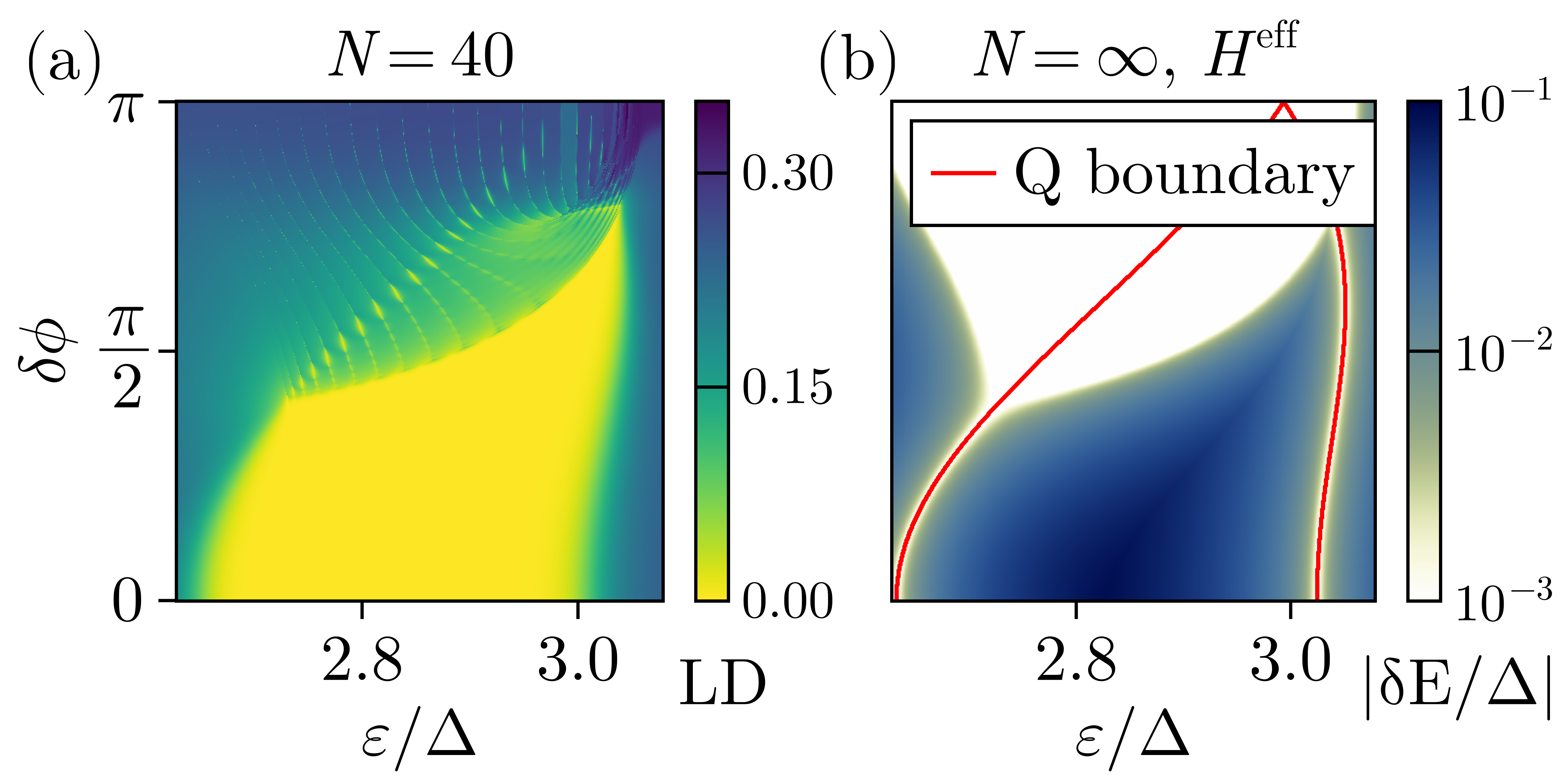}
    \caption{Phase diagrams for finite and infinite systems. (a) In an open system with 40 sites $\mathrm{LD}$ shows clear phase boundaries. (b) In an infinite, periodic system, with the second order effective Hamiltonian, there is a topological phase and a trivial phase separated by the red contour where the topological invariant changes sign and the gap closes at $k=0$ or $k=\pi$. The heatmap shows the energy gap which also reveals an extended region where the gap closes at a finite momentum.}
    \label{fig:phase_diagrams}
\end{figure}

Fig.\,\ref{fig:phase_diagrams}(a) shows the tuning diagram for a 40-dot chain, where a phase diagram starts taking shape. For long enough chains, any point in the topological phase has good protection and energy degeneracy and there is no need to fine-tune to a sweet spot. The patterns seen in Fig.\,\ref{fig:phase_diagrams}(a) can be matched with the true phase diagram of the infinite system in \ref{fig:phase_diagrams}(b). Since the second order effective model captures the physics well, we use it to determine a topological invariant $Q$ (see App.\,\ref{app:topological_invariant}) that signifies when the gap closes at $k=0$ or $k=\pi$. It can take on the values $+1$ (trivial phase) or $-1$ (topological phase) and the boundary between these is plotted as the red contour in Fig.\,\ref{fig:phase_diagrams}(b). Due to the breaking of time reversal symmetry, there is also a gapless phase where the gap closes at finite momentum \cite{degottardiMajoranaFermionsSuperconducting2013}, which can be seen in the heatmap of Fig.\,\ref{fig:phase_diagrams}(b) that shows the energy gap of the infinite, periodic system.

\section{Summary and conclusions}
In this work, we have introduced the local distinguishability of ground states as a quality measure of near-topological phases in finite and (possibly) strongly interacting systems, and applied this measure to quantum-dot-based Kitaev chains of varying length. It has a clear experimental signature in that it bounds the energy splitting of any local perturbation. Conversely, local measurements provide lower bounds on $\mathrm{LD}$, but determining the actual value requires in general performing \textit{all} local measurements. 

As expected, the local distinguishability decreases exponentially for long chains indicating a transition to a true topological phase, but for short chains the pattern is different. In particular, we have shown that the three-dot case can actually be worse than the two-dot case. The explanation for this result is that the long-range nature of the effective Hamiltonian introduces a trade-off between separated MBSs and the energy degeneracy between the ground states. This has consequences for protocols aiming to demonstrate the non-abelian nature of MBSs, which rely on both energy degeneracy and the MBSs not overlapping \cite{tsintzisMajoranaQubitsNonAbelian2024}. It would be an interesting direction for future works to investigate if non-abelian protocols can be designed that only require the ground state energy splitting to be \textit{stable}, not zero. On a more general level, much remains to be understood concerning the topological-like protection and nonabelian properties of strongly interacting, finite systems.

\begin{acknowledgments}
We acknowledge stimulating discussions with Chun-Xiao Liu, William Samuelson, Michael Wimmer and Mert Bozkurt and funding from the European Research Council (ERC) under the European Unions Horizon 2020 research and innovation
programme under Grant Agreement No. 856526, the
Swedish Research Council under Grant Agreement No.
2020-03412, and NanoLund.
\end{acknowledgments}


\appendix

\section{Local Distinguishability and MBSs}\label{app:LD}
Here, we show that $\norm{\delta\rho_R} = 0$ if there exists a Majorana operator in the complement. If there exists a Hermitian operator $\gamma = I_R \otimes \gamma_{R^\complement}$ such that $\ket{e} = \gamma\ket{o}$ and $\gamma_{R^\complement}^2 = I$, then by the cyclic property of the partial trace 
\begin{equation}
    \Tr_{R^\complement}\dyad{e} = \Tr_{R^\complement} \big[ \gamma_{R^\complement} \dyad{o} \gamma_{R^\complement} \big] = \Tr_{R^\complement}\dyad{o}
\end{equation}
and therefore $\delta\rho_R = 0$. This is a necessary condition for the existence of MBSs outside of $R$, but not sufficient. 

If there are several such operators, then $\delta\rho_R$ can only be non-zero in regions that include all of them, as it vanishes if any one of them is outside $R$.

\section{More on branches}\label{app:branches}
\begin{figure}[t!]
    \centering
    \includegraphics[width=1\linewidth]{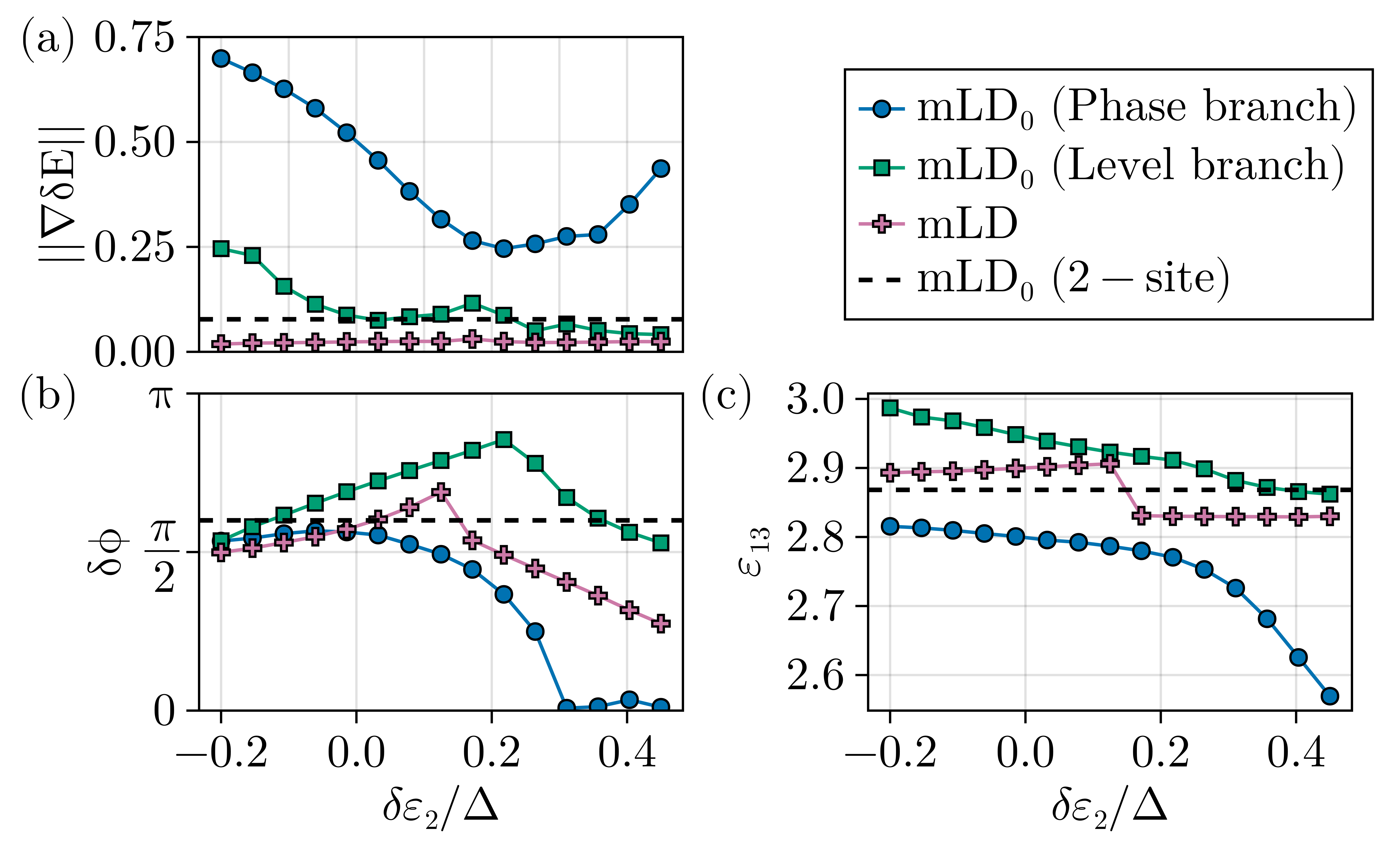}
    \caption{Similar to Fig.\,\ref{fig:3site_inhomogeneous}, we follow different sweet spots as a function of the detuning $\delta\varepsilon_2$. (a) norm of the gradient of $\delta E$. (b) $\delta\phi$ at the sweet spot. (c) $\varepsilon$ at the sweet spot.}
    \label{fig:sweet_spot_branches_appendix}
\end{figure}

In Fig.\,\ref{fig:sweet_spot_branches_appendix} some additional plots for the three-dot chain as a function of the detuning of the middle dot, complementing Fig\,\ref{fig:3site_inhomogeneous}. 

We plot the gradient of the ground state energy splitting $\norm{\boldsymbol{\nabla}\delta E}$ in Fig.\,\ref{fig:sweet_spot_branches_appendix}(a), where
$\boldsymbol{\nabla} \equiv \begin{bmatrix} \frac{\partial}{\partial\varepsilon} & \frac{\partial}{\partial\delta\phi} \end{bmatrix}$. Note that we still optimize for a minimal $\mathrm{LD}$, not for minimal gradient. The gradient differs somewhat from the $\mathrm{LD}$ since it is sensitive to details such as how the Hamiltonian is parameterized. In other words, it also includes the effect of type 1 and 2 protection as explained in Sec.\,\ref{sec:protection_types}. The location of the sweet spots is shown in Figures \ref{fig:sweet_spot_branches_appendix}(b) and (c), where we see that the $\mathrm{mLD}_0$ points on the level and phase branch can be followed continuously while the $\mathrm{mLD}$ has a discontinuous jump. 

\section{Interactions}\label{app:interactions}
\begin{figure}[t!]
    \centering
    \includegraphics[width=1\linewidth]{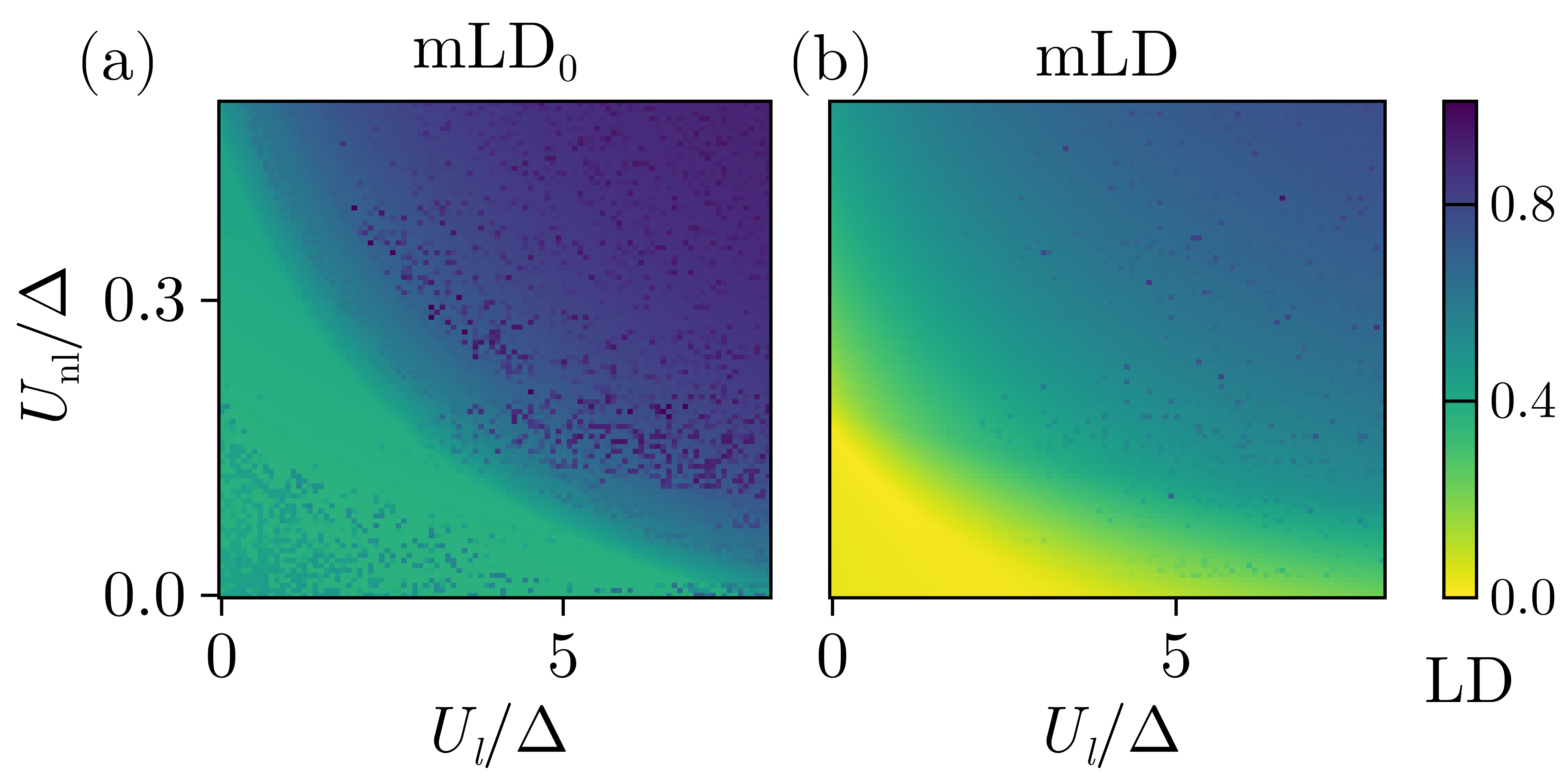}
    \caption{$\mathrm{LD}$ at $\mathrm{mLD}_0$ (a) and at $\mathrm{mLD}$ (b) as a function of the strength of local $U_l$ and non-local $U_{\mathrm{nl}}$ Coulomb interaction. This is similar to the two-dot chain studied in Ref.\,\cite{samuelsonMinimalQuantumDot2024}. The system can only tolerate a certain amount of interaction before strongly degrading the quality of the sweet spot.}
    \label{fig:interactions}
\end{figure}

\begin{figure}
    \centering
    \includegraphics[width=1\linewidth]{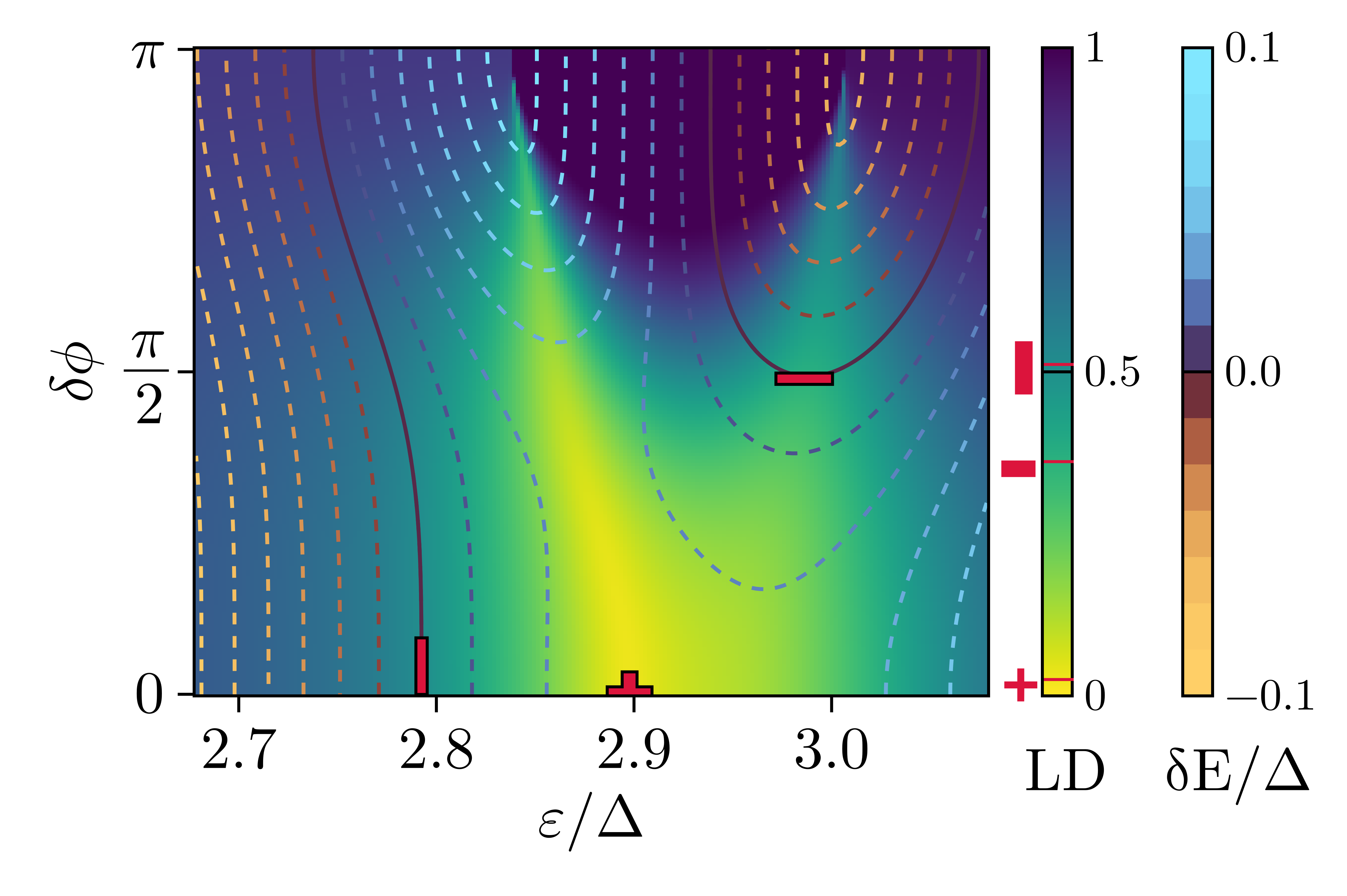}
    \caption{LD as a function of $\delta \phi$ and $\varepsilon$. Same as Fig.\,\ref{fig:tuning_sweet_spot} but with finite interactions where $U_l = 3\Delta, U_{\mathrm{nl}} = \Delta/100$.}
    \label{fig:tuning_interactions}
\end{figure}

For the main conclusions of this article, we found that that interactions did not play a central role, and excluded them for simplicity. In Ref.\,\cite{samuelsonMinimalQuantumDot2024}, the role of interactions were studied in the two-dot system. There, the main conclusion was that if they are sufficiently small, they only slightly affect the quality of the sweet spot, while if they are too large, they may prohibit the tuning to the sweet spot. For the three-dot chain, we find qualitatively the same result, see Fig.\,\ref{fig:interactions}. In Fig.\,\ref{fig:tuning_interactions} we show how the tuning plot changes when interactions are included.

\section{\label{app:perturbations}Effective Hamiltonian}
In terms of the Bogoliubov quasiparticles where $H_\mathrm{QD}$ is diagonal, the Hamiltonians takes the form
\begin{align}
    H^\text{QD} = &\sum_n \varepsilon_a a_n^\dagger a_n  + \varepsilon_b b_n^\dagger b_n \\
    \begin{split}
    H^C = &\sum_n t_{aa} a_n^{\dagger } a_{n+1} +\Delta_{aa} a_n a_{n+1} + \\
    &+t_{ab}a_n^{\dagger } b_{n+1} + \Delta_{ab} a_n b_{n+1} -\Delta_{ba} a_{n+1} b_n\\
    &+t_{ba}b_n^{\dagger } a_{n+1}+\Delta_{bb} b_n b_{n+1}+t_{bb} b_n^{\dagger } b_n + \mathrm{h.c},
    \end{split}
\end{align}
where
\begin{subequations}
\begin{align}
\varepsilon_a &= \sqrt{\varepsilon^2 + \Delta^2} - V_Z \\
\varepsilon_b &= \sqrt{\varepsilon^2 + \Delta^2} + V_Z \\
t_{aa} &= -\frac{t \varepsilon  \cos \left(\frac{\delta \phi }{2}\right)}{\sqrt{\Delta ^2+\varepsilon ^2}}+i t \sin \left(\frac{\delta \phi }{2}\right) \\
t_{ab} &= -\frac{\Delta  t \cos \left(\frac{\delta \phi }{2}\right)}{\sqrt{\Delta ^2+\varepsilon ^2}} \\
\Delta_{aa} &= -\frac{\Delta t_\text{so} \cos \left(\frac{\delta \phi }{2}\right)}{\sqrt{\Delta ^2+\varepsilon ^2}} \\
\Delta_{ab} &= \frac{t_\text{so} \varepsilon  \cos \left(\frac{\delta \phi }{2}\right)}{\sqrt{\Delta ^2+\varepsilon ^2}}+i t_\text{so} \sin \left(\frac{\delta \phi }{2}\right)\\
t_{ba} &= t_{ab} \\
t_{bb} &= t_{aa}^* \\
\Delta_{bb} &= -\Delta_{aa}\\
\Delta_{ba} &= \Delta_{ab}^*
\end{align}
\end{subequations}

The first order term in the effective Hamiltonian is the projection onto states with no $b$-fermions,
\begin{equation}
    H_1^\text{eff} = \sum_n \varepsilon_a a_n^\dagger a_n+ (t_{aa} a_n^{\dagger } a_{n+1} +\Delta_{aa} a_n a_{n+1} + \mathrm{h.c}).
\end{equation}
The second order term is
\begin{multline}
    H_2^\text{eff} = \sum_n \varepsilon_2 a_n^\dagger a_n + t_2 a_n^\dagger a_{n+2} + \Delta_2 a^\dagger_{n}a^\dagger_{n+2} + \mathrm{h.c}
\end{multline}
where 
\begin{align}
   \varepsilon_{2} &= 2(|\Delta_{ab}|^2  - |t_{ab}|^2) / \varepsilon_b \\
    t_{2} &= (-\Delta_{ab}'^2  - t_{ab}^2) / \varepsilon_b \\
    \Delta_{2} &= 2\Re{\Delta_{ab}  t_{ab}} / \varepsilon_b 
\end{align}
It includes the effect of next nearest neighbour hoppings and pairings.

\section{Topological invariant}\label{app:topological_invariant}
The topological invariant for the second order effective model (which is in class \textbf{D}) signifies when the gap closes at $k=0$ or $k=\pi$ can be characterized by 
\begin{equation} 
    Q \equiv \operatorname{sign}(\operatorname{Pf}(H_0)\operatorname{Pf}(H_\pi)).
\end{equation}
In the second order effective model, this is simply
\begin{equation} 
    Q \equiv \operatorname{sign}((2\Re{t_1+t_2}+ \varepsilon)(2\Re{-t_1+t_2}+ \varepsilon).
\end{equation}
However, it should be noted that the gap can also close a finite momentum due to the breaking of time-reversal symmetry \cite{degottardiMajoranaFermionsSuperconducting2013}.

\bibliography{LongerPoorMansArticle}

\end{document}